\documentclass[twocolumn,superscriptaddress,showpacs,amsfonts,amsmath,amssymb]{revtex4-1}

\usepackage{epsfig}
\usepackage{amsmath}
\usepackage{amssymb}
\usepackage{amsfonts}
\usepackage{mathptmx}
\usepackage{dcolumn}
\usepackage{eucal}
\usepackage{bm}
\usepackage{color}
\usepackage[colorlinks,linkcolor=blue,citecolor=blue]{hyperref}
\usepackage{epstopdf}
\usepackage{bbold}
\usepackage{url}
\usepackage{siunitx}
\usepackage{dsfont}

\usepackage{soul}%For striking out text

\def\eps{\varepsilon}

\def\bk{{\bf k}}

\def\up{\uparrow}
\def\dw{\downarrow}
\def\be{\begin{equation}}
\def\ee{\end{equation}}
\def\bea{\begin{eqnarray}}
\def\eea{\end{eqnarray}}
\def\ba{\begin{array}{l l}}
\def\ea{\end{array}}

\begin{document}

\title{Unveiling Mechanisms of Electric Field Effects on Superconductors by Magnetic Field Response}

\author{Lennart Bours}
\affiliation{NEST, Istituto Nanoscienze--CNR and Scuola Normale Superiore, Piazza San Silvestro 12, 56127 Pisa, Italy}
\author{Maria Teresa Mercaldo}
\affiliation{Dipartimento di Fisica ``E. R. Caianiello'', Universit\`a di Salerno, IT-84084 Fisciano (SA), Italy}
\author{Mario Cuoco}
\affiliation{SPIN-CNR,  IT-84084  Fisciano  (SA),  Italy}
\affiliation{Dipartimento di Fisica ``E. R. Caianiello'', Universit\`a di Salerno, IT-84084 Fisciano (SA), Italy}
\author{Elia Strambini}
\affiliation{NEST, Istituto Nanoscienze--CNR and Scuola Normale Superiore, Piazza San Silvestro 12, 56127 Pisa, Italy}
\author{Francesco Giazotto}
\affiliation{NEST, Istituto Nanoscienze--CNR and Scuola Normale Superiore, Piazza San Silvestro 12, 56127 Pisa, Italy}
\email{francesco.giazotto@sns.it}

\begin{abstract}
We demonstrate that superconducting aluminium nano-bridges can be driven into a state with complete suppression of the critical supercurrent via electrostatic gating. Probing both in- and out-of-plane magnetic field responses in the presence of electrostatic gating can unveil the mechanisms that primarily cause the superconducting electric field effects. Remarkably, we find that a magnetic field, independently of its orientation, has only a weak influence on the critical electric field that identifies the transition from the superconducting state to a phase with vanishing critical supercurrent. This observation points to the absence of a direct coupling between the electric field and the amplitude of the superconducting order parameter or $2\pi$-phase slips via vortex generation.
The magnetic field effect observed in the presence of electrostatic gating is described within a microscopic model where a spatially uniform inter-band $\pi$-phase is stabilized by the electric field. Such an intrinsic superconducting phase rearrangement can account for the suppression of the supercurrent, as well as for the weak dependence of the critical magnetic fields on the electric field. 
\end{abstract}
\maketitle

\section{Introduction}
Recently, it has been shown that the superconducting (SC) properties of metallic Bardeen-Cooper-Schrieffer (BCS) superconductors can be influenced via electrostatic gating~\cite{Paolucci2019c}. The most striking effect: reduction and suppression of the critical supercurrent, has been broadly demonstrated in metallic nanowires\cite{DeSimoni2018, Ritter2020,Alegria2020} and Dayem bridges\cite{Likharev1979,Paolucci2018,Paolucci2019, Puglia2020} made of titanium, titanium nitrate, aluminum, niobium and vanadium, as well as in aluminum--copper--aluminum Josephson junctions~\cite{DeSimoni2019}. Moreover, recent experiments have probed the effect of electrostatic gating on the SC-phase in a SQUID~\cite{Paolucci2019b}, and on the nature of the switching current distributions in gated titanium Dayem bridges~\cite{Puglia2019}.

While these observations clearly indicate that the electric field can suppress the supercurrent, whether and how it acts on the amplitude or the phase of the SC order parameter are questions so far unanswered. To develop a deeper insight into this fundamental problem we investigate how the SC state is modified by the simultaneous presence of electric and magnetic fields. In this context, probing both the in-, and out-of-plane magnetic fields ($B_\text{Y}$ and $B_\text{Z}$, respectively) is particularly useful as the two orientations affect the SC thin films via very different mechanisms~\cite{Fulde1973}. In thin films $B_\text{Z}$ generally leads to screening currents and a spatially varying order parameter, marked by $2\pi$-phase slips, as flux vortices penetrate the sample. $B_\text{Y}$ on the other hand, ideally affects the pairing amplitude homogeneously via electron spin paramagnetism, inducing pair breaking and spin polarization~\cite{Chandrasekhar1962,Adams2017}. Thus, the search for magneto-electric cross-talking effects in superconducting thin films can provide indications and constraints on the quantum states at superconductivity breakdown, and reveal the origin of the unexpected coupling between the electric field and the SC phase and/or pairing amplitude.  

We demonstrate that SC Al nano-bridges can be electrically driven into a state with complete suppression of the critical supercurrent, and investigate their response to both in- and out-of-plane magnetic fields at various temperatures. While the effects of an out-of-plane magnetic field in combination with an electric field have been measured in long Ti nanowires~\cite{DeSimoni2018} at low temperature, this is the first time the combined electric and magnetic field response of a superconducting Dayem bridge has been characterized. Also, in contrast to the previous work, the effect is investigated at several temperatures, to map the whole temperature dependence of these effects as well. 

Remarkably, we find that the magnetic field has only a weak influence on the electric field effect in the SC bridges. Moreover, this phenomenology is starkly independent on the magnetic field orientation, despite the very different interactions between SC thin films and in-, and out-of-plane magnetic fields. These findings suggest the absence of a direct electric coupling between the electric field and the amplitude of the SC order parameter, or 2$\pi$ phase slips generated by vortices. Both cases would have manifested with a significant variation of the SC/normal (N) critical boundaries in the presence of magnetic fields. 

Our observations appear consistent with a recently proposed and here further developed model in which the surface electric field is a source of inversion-symmetry breaking interactions that strongly affects the orbital polarization only at the surface layers of a multi-band superconducting thin film~\cite{mercaldo2019}. This results in an electric-field-driven phase transition into a mixed superconducting state where the relative SC phases between different bands are shifted by $\pi$. This state, apart from naturally yielding a suppression of the supercurrent, is proven to be hardly influenced by the applied magnetic field, as shown in the phase diagram, thus capturing the main experimental findings. 

In addition to the fundamental aspects discussed above, the full suppression of the supercurrent is for the first time demonstrated in our Al-based devices (previously only a 35\% reduction was achieved for Al wires~\cite{DeSimoni2018}). This is noteworthy as aluminum has several useful and impactful properties. Indeed, it is easy to evaporate, has a self limiting oxide layer, and in situ its oxide layers can be reliably controlled by make fine tuned tunneling barriers, which makes our results significant from the technological point of view. Moreover, considering the broad application of Al-based thin films as SC qubits~\cite{Welander2013}, Josephson devices~\cite{Makhlin2001,Pekola2013,Xiang2013}, photon detectors~\cite{Holland1999,Schmidt2011} and bolometers~\cite{Klapwijk2017}, one can envision a new generation of SC electronics that can fully exploit the demonstrated SC electric field effects.
%Another motivation for investigating the Al refers to its electronic %character with $p$- bands at the Fermi level. This is a relevant case from %the point of view of evaluating the applicability of the proposed %microscopic mechanism and the role of the material electronic structure. %Indeed, as our model description indicates that for the electrical %tunability one needs states with non-vanishing orbital angular momentum, %it is interesting to confirm this microscopic hypothesis for %superconductors with $p$-bands and compare it with other already tested %cases having $d$-bands (e.g. Ti, V and Nb).

The paper is organized as follows. In Sect. II we present the experimental methods and the employed theoretical model including aspects for the computation of the phase diagram.
Sect. III is devoted to the results concerning the evolution of the critical supercurrent as a function of the applied magnetic and electric fields, and the corresponding theoretical analysis of the phase diagram.
Sec. IV is assigned to discussion and concluding remarks.   

\begin{figure}
\centering
\includegraphics[width=\columnwidth]{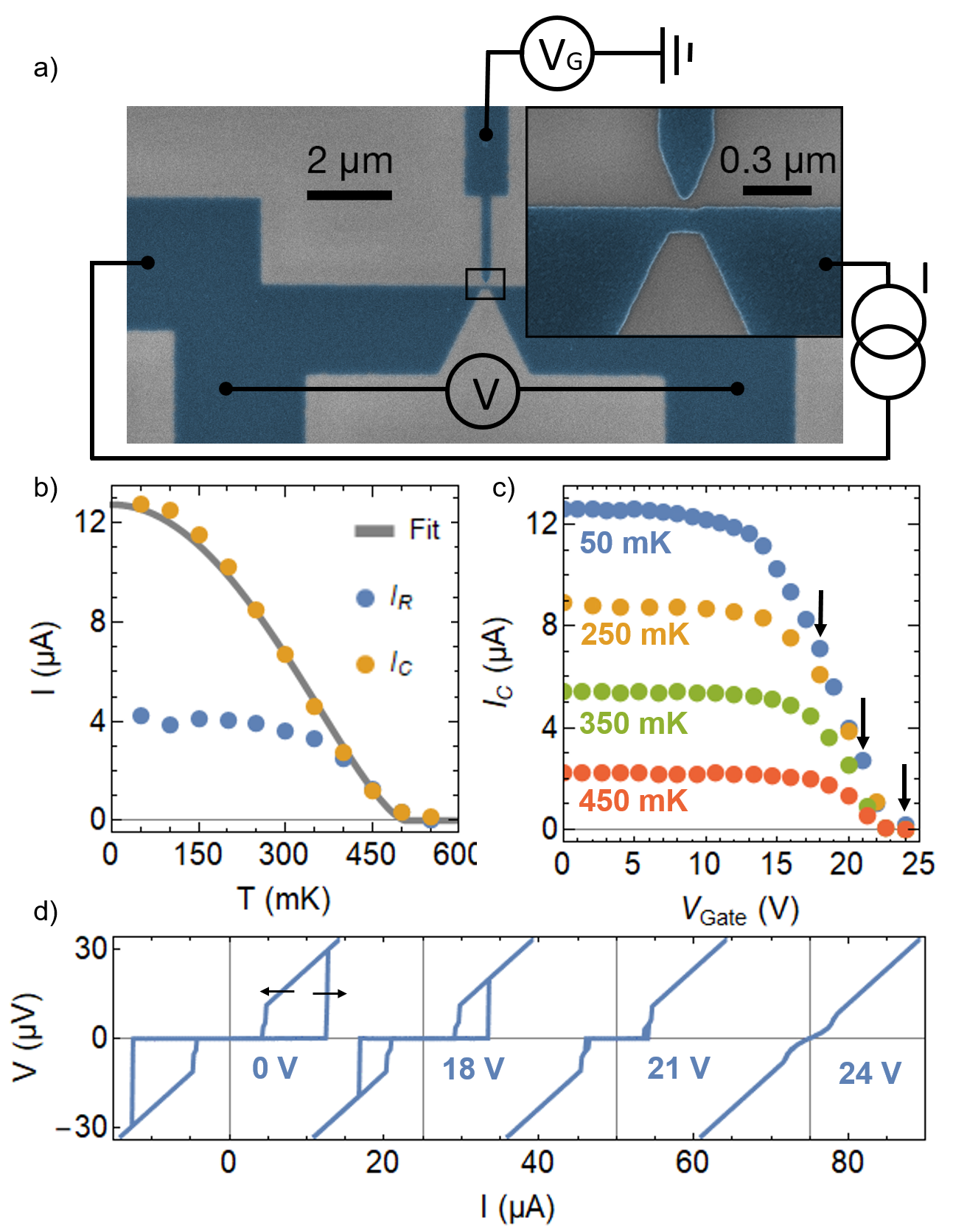}
\caption{a) False color SEM image of a typical device. Inset: close up of the region indicated by the black square showing the Dayem bridge and gate electrode. 
b) Critical current $I_\text{C}$ and retrapping current $I_\text{R}$ versus temperature. $I_\text{C}$ follows the typical BCS evolution (gray line). 
c) $I_\text{C}$ versus gate voltage $V_{\text{Gate}}$ at four different temperatures.
d) Voltage drop across the bridge versus bias current, for four values of $V_{\text{Gate}}$ at 50 mK. Arrows indicate sweep direction, and the curves are horizontally offset for clarity. The black arrows in c) indicate the curves with $V_\text{Gate} = $ 18, 21 and 24 V.
}	
\label{fig:device}
\end{figure}

\section{Methods}

In this section we discuss the methodologies for the fabrication and design of the device together with the main aspects related to the transport measurements. Moreover, we describe the model employed for the investigation of the superconducting phase in the presence of both electric and magnetic fields. 

\subsection{Experimental device: fabrication and transport}
The samples were fabricated in a single step employing electron-beam lithography to pattern a resist mask on a sapphire substrate, see Fig.~\ref{fig:device}a. 3 nm of titanium was deposited at 1 \AA/s (to improve adhesion), after which 14 nm of aluminium was deposited at 2.5 \AA/s, in an electron beam evaporator with a base pressure of $\approx 10^{-11}$ torr. The Dayem bridge is approximately 120 nm wide, 100 nm long, and has a normal-state resistance $R_\text{N} \approx 25 \,\Omega$. Gate-bridge separation is about 30 nm, and the leads on either side of the bridge are \SI{2}{\micro\metre} wide. Resistance versus temperature measurements, performed using a \SI{3}{\micro\volt} square wave excitation indicate a critical temperature $T_\text{C} \approx 600$ mK, and a transition width of $\approx$ 60 mK. 

The critical current measurements were performed in a He-3 He-4 dilution refrigerator at temperatures ranging from 50 to 600 mK, using a standard four wire set-up, biasing with a current. The DC lines are fitted with low pass and $\pi$ filters. The voltage drop was amplified using a room temperature differential pre-amplifier, while the gate voltage was supplied by a low noise sourcemeter. To determine the critical current $I_\text{C}$, current-voltage $I-V$ measurements were repeated 30 to 50 times.

The leakage current between the gate and device was carefully measured by applying a voltage to the gate in the usual manner, and amplifying the current flowing into the device using a room temperature current amplifier over a long period of time. At $V_\text{Gate} = 25$ V, the leakage current $I \approx 7 * 10^{-11}$ A, giving a gate-device resistance of $R \approx 0.63$ T$\Omega$. This is of the same order of magnitude as reported in previous works.\cite{Paolucci2019c}

Using the BCS relation, we find that $\Delta_0 = 1.764~ k_\text{B} T_\text{C} = $ \SI{91}{\micro\electronvolt} ($k_\text{B}$ being the Boltzmann constant). Here, $T_\text{C}$ is relatively low for Al, likely due to an inverse proximity effect from the Ti layer. Via the conductivity $\sigma$, $\Delta_0$ and the magnetic permeability of the vacuum $\mu_0$, we estimate the London penetration depth $\lambda_\text{L} = \sqrt{\hbar / \mu_0 \pi \sigma \Delta_0} \approx 100$ nm, and the superconducting coherence length $\xi_0 = \sqrt{\hbar \sigma / N_\text{F} e^2 \Delta_0} \approx 170$ nm. Here, we take the electron density at the Fermi energy of aluminum to be $N_\text{F} = 2.15 \cdot 10^{47}~ \text{J}^{-1} \text{m}^{-3}$~\cite{Anthore2003,Maisi2013,Ligato2017}. Although the critical temperature of Ti Dayem bridges is similar to that of the Al, we point out that the Ti is not contributing to the observed effects. This can be deduced by the observed critical magnetic fields for the Al which are very different from those of the Ti. 

\subsection{Model and computation}
In order to capture the effects of magnetic and electric fields we introduce a microscopic model to simulate multiband superconductivity with conventional $s$-wave spin-singlet pairing for a slab geometry with $n_z$ layers~\cite{mercaldo2019}. 
The electric field, $E_s$, on the surface is parallel to $\hat{z}$, and thus it can be described by a potential $V_{s}=-E_{s} z$. 
%Following the approach already applied to derive the surface orbital Rashba coupling \cite{Park2011,Park2012,Kim2013}, the matrix elements of $V_s$ in the Bloch basis lead to an intra- ($\alpha_{OR})$ and inter-layer ($\lambda$) inversion asymmetric interactions, whose amplitude is proportional to $E_s$ while the relative ratio depends on the inter-atomic distances and distortions at the surface (see Appendix A for details).
Following the customary approach we derive the surface orbital Rashba coupling~\cite{Park2011,Park2012,Kim2013}. The matrix elements of $V_s$ lead to an intra- ($\alpha_{OR})$ and inter-layer ($\lambda$) inversion asymmetric interactions (in the Bloch basis), whose amplitude is proportional to $E_s$ while the relative ratio depends on the inter-atomic distances and distortions at the surface (see Appendix A for details).
In this context, the effect of the electric field in driving an orbital polarization is particularly relevant for materials with $p$- or $d$-orbitals at each atomic site. Hence, the case of aluminum is included in the proposed modelling because $p$-bands contribute to the Fermi level. 
%In this subsbace, one can encode the effects of inversion symmetry breaking by introducing orbital dependent asymmetric couplings at the surface layers. 

Here, for convenience and clarity we indicate as $(a,b,c)$ the three orbitals on each atom that we employ for building up the tight-binding model and can refer to either $p$-bands or a subspace of the $d$-manifold. Then, assuming translational invariance in the $xy$ planes, we introduce the creation $d^\dagger_{\alpha,\sigma}(\bk,i_z)$ and annihilation $d_{\alpha,\sigma}(\bk,i_z)$ operators with momentum $\bk$, spin ($\sigma=[\up,\dw]$), orbital ($\alpha=(a,b,c$)), and layer $i_z$, to construct a spinorial basis
$\Psi^\dagger(\bk,i_z)=(\Psi_{\up}^\dagger(\bk,i_z), \Psi_{\dw}(-\bk,i_z))$ with
$\Psi_{\sigma}^\dagger(\bk,i_z)=(d^\dagger_{a,\sigma}(\bk,i_z),d^\dagger_{b,\sigma}(\bk,i_z),d^\dagger_{c,\sigma}(\bk,i_z))$. In this representation, the complete Hamiltonian can be expressed in a compact way as:
\begin{eqnarray}
\mathcal{H}= \frac{1}{N} \sum_{\bk,i_z,j_z} \Psi^{\dagger}(\bk,i_z) \hat{H}(\bk) \Psi(\bk,j_z) \,,
\end{eqnarray}
\noindent with 
\begin{eqnarray}
\hat{H}(\bk)=\hat{H}_{SC}+\hat{H}_{ISB}+\hat{H}_M
\end{eqnarray}
\noindent where $\hat{H}_{SC}$ is the superconducting part related with the multilayered configuration assuming an intra-orbital singlet pairing, 
\begin{eqnarray}
\hat{H}_{SC}=&&\sum_{\alpha} [\tau_z \eps_\alpha(\bk) + \Delta_\alpha(i_z) \tau_x]\otimes(\hat{L}^2 -2\hat{L}^2_\alpha)] \delta(i_z,j_z) + \nonumber \\
&& + t_{\perp,\alpha} \tau_z \otimes (\hat{L}^2 - 2 \hat{L}^2_\alpha) \delta(i_z,j_z\pm 1) \, ,
\end{eqnarray}

\noindent the term $\hat{H}_{ISB}$ arises from the inversion symmetry breaking at the surface of the superconducting thin film due to the presence of the electric field and is expressed as
\begin{eqnarray}
&&\hat{H}_{ISB}=\alpha_{OR} \tau_z \otimes (\sin k_y \hat{L}_x -\sin k_x \hat{L}_y) [\delta(i_z,j_z)(\delta(i_z,1) +  \nonumber \\ && \delta(i_z,n_z)]  +  \lambda \tau_0 \otimes (\hat{L}_x+\hat{L}_y) [\delta(i_z,1)\delta(j_z,2)-\delta(i_z,2)\delta(j_z,1)+\nonumber \\ 
&& \delta(i_z,n_z)\delta(j_z,n_z-1)-\delta(i_z,n_z-1)\delta(j_z,n_z)] \, ,
\end{eqnarray} 
\noindent and finally the Zeeman term is given by
\begin{eqnarray}
&&\hat{H}_{M}=B\, \sum_{\alpha} \tau_0 \otimes(\hat{L}^2 -2\hat{L}^2_\alpha) \delta(i_z,j_z)
\end{eqnarray} 
Here, the orbital angular momentum operators $\hat{L}$ have components 
$
\hat{L}_x=\begin{bmatrix}
0 & 0 & 0 \\ 
0 & 0 & i \\ 
0 & -i & 0%
\end{bmatrix}, 
\hat{L}_y=\begin{bmatrix}
0 & 0 & -i \\ 
0 & 0 & 0 \\ 
i & 0 & 0%
\end{bmatrix}, \hat{L}_z=\begin{bmatrix}
0 & -i & 0 \\ 
i & 0 & 0 \\ 
0 & 0 & 0%
\end{bmatrix}$
within the ($a,b,c$) subspace, $\tau_i$ ($i=x,y,z$) are the Pauli matrices for the electron-hole sector, and $\delta_{i,j}$ the Kronecker delta function. 
The kinetic energy for the in-plane electron itinerancy is due to the symmetry allowed \cite{Slater1954} nearest neighbor hopping, thus, one has that $\eps_a(\bk)=-2 t_{||} [\cos(k_x)+ \eta \cos(k_y)]$, $\eps_b(\bk)=-2 t_{||} [\eta \cos(k_x)+\cos(k_y)]$, and $\eps_c(\bk)=-2 t_{||} [\cos(k_x)+ \cos(k_y)]$, with $\eta$ being a term that takes into account deviations from the ideal cubic symmetry. 
%The role of inter-orbital hopping that are activated by distortions has been explicitly evaluated and the %outcome is not affected by these terms.
We assume that the layer dependent spin-singlet OP is non-vanishing only for electrons belonging to the same band and it is expressed as $\Delta_\alpha(i_z)=\frac{1}{N} \sum_{\bk} g\,\langle d_{\alpha,\uparrow}(\bk,i_z) d_{\alpha,\downarrow}(-\bk,i_z)  \rangle$ with $\langle ...\rangle$ being the expectation value on the ground state. 
Here, $N=n_x \times n_y$ sets the dimension of the layer in terms of the linear lengths $n_x$ and $n_y$, while we assume translation invariance in the $xy$-plane and $n_z$ layers along the $z-$axis. We notice that in the present analysis the pairing strength $g$ is not modified by the electric field. This is physically consistent with the fact that due to screening effects the electric field cannot induce an inversion asymmetric potential inside the thin film beyond the Thomas-Fermi length.
The study is then conducted by determining the superconducting OPs corresponding to the minimum of the free energy.
The planar hopping is the energy unit, $t_{||}=t$, while the interlayer one is orbital independent, i.e. $t_{\perp,\alpha}=t_{\perp}$, and the pairing coupling is $g=2\,t$. A change in the pairing interaction does not qualtitatively alter the phase diagram \cite{mercaldo2019}. 

\begin{figure*}[t!]
\centering
\includegraphics[width=\textwidth]{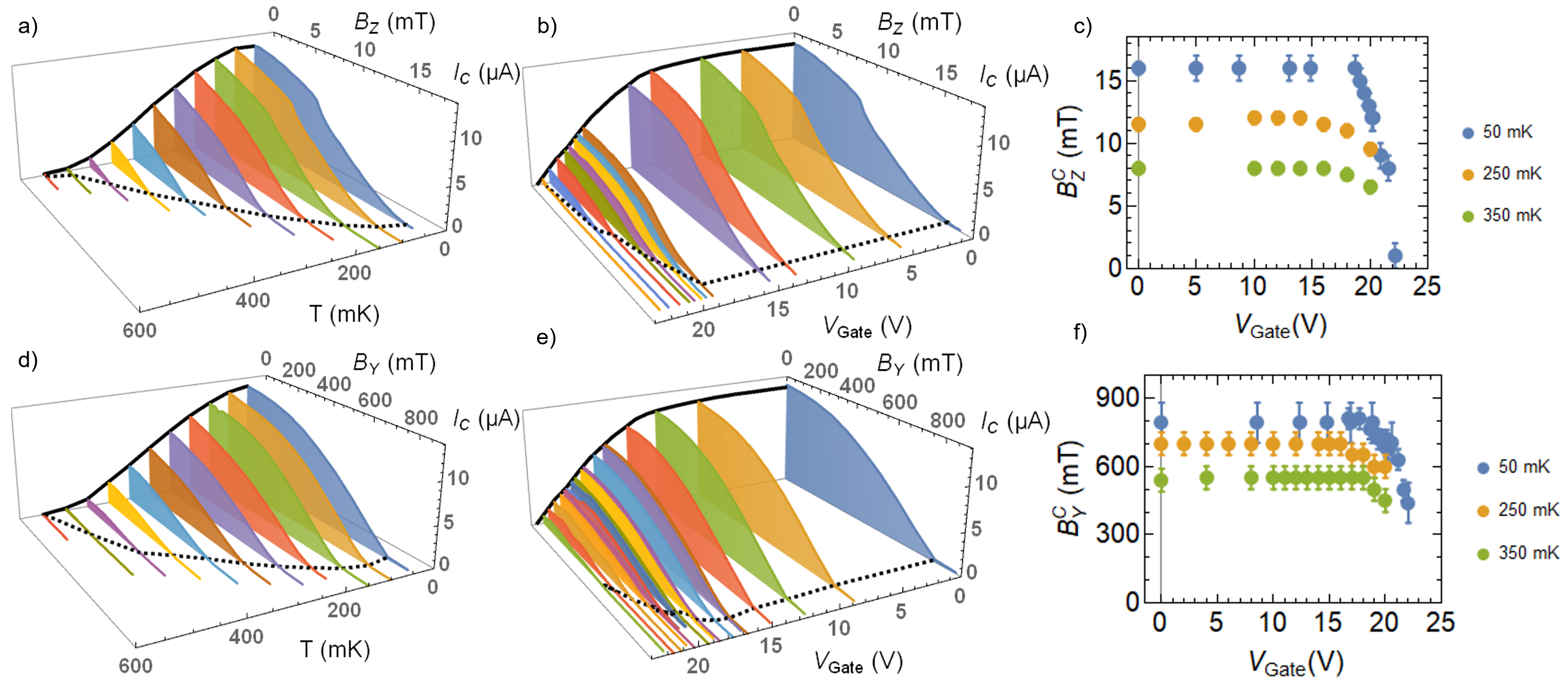}
\caption{a) 3D plot of $I_\text{C}$ as a function of the out-of-plane magnetic field $B_\text{Z}$ and temperature $T$. The full black line shows $I_\text{C} (T)$ at zero field, the dashed black line shows the critical field $B_\text{Z}^\text{C}$ versus temperature.
b) 3D plot of $I_\text{C}$ as a function of the out-of-plane magnetic field $B_\text{Z}$ and the gate voltage at 50 mK. The full black line indicates $I_\text{C} (V_\text{Gate})$ at zero field, the dashed black line indicates the critical field $B_\text{Z}^\text{C}$.
c) Critical $B_\text{Z}^\text{C}$ versus $V_\text{Gate}$ at $T=$ 50, 250 and 350 mK. The error bars indicate the resolution in $B_\text{Z}$.
d) 3D plot of $I_\text{C}$ as a function of the in-plane magnetic field $B_\text{Y}$ and temperature $T$. The full black line shows $I_\text{C} (T)$ at zero field, the dashed black line shows the critical field $B_\text{Y}^\text{C}$ versus temperature.
e) 3D plot of $I_\text{C}$ as a function of the in-plane magnetic field $B_\text{Y}$ and the gate voltage at 50 mK. The full black line indicates $I_\text{C} (V_\text{Gate})$ at zero field, the dashed black line indicates the critical $B_\text{Y}^\text{C}$.
f) Critical field $B_\text{Y}^\text{C}$ versus $V_\text{Gate}$ at $T =$ 50, 250 and 350 mK. The error bars indicate the resolution in $B_\text{Y}$.}
\label{fig:BEverything}
\end{figure*}

\section{Results}

In this section we present the experimental and theoretical phase diagrams in terms of applied magnetic fields, electrostatic gating and temperature.

\subsection{Electric field vs in- and out-of-plane magnetic field}
 
We start by discussing the behavior of the supercurrent at zero applied magnetic field by varying the amplitude of the electrostatic field and the temperature.
Fig.~\ref{fig:device}b shows the critical and retrapping currents versus temperature. At the base temperature of 50 mK, the critical current $I_\text{C} \approx \SI{12.8}{\micro\ampere}$. The evolution of $I_\text{C}$ as a function of temperature follows the conventional Bardeen's profile \cite{Bardeen1962,Song1972,Khlebnikov2017} $I_\text{C} \cong I_\text{C}^0 [1-(\frac{T}{T_\text{C}})^2]^{3/2}$. The IV characteristics show a considerable hysteresis at low temperature (see the blue dots in Fig.~\ref{fig:device}b and the lines in Fig.~\ref{fig:device}d), with a retrapping current $I_\text{R} \approx  \SI{4.2}{\micro\ampere}$ at $T=50$ mK. The hysteresis is likely thermal in origin~\cite{Skocpol1974,Courtois2008,Hazra2015}, and it disappears when $T>400$ mK, which is consistent with an enhanced thermalization mediated by phonon coupling.

As in similar experiments~\cite{DeSimoni2018,Paolucci2018,DeSimoni2019,Paolucci2019b,Puglia2019,Paolucci2019,Paolucci2019c}, the critical current can be reduced, up to complete suppression at the critical gate voltage $V_\text{Gate}^{C} \approx 23$ V. This is shown in Fig.~\ref{fig:device}c, for several temperatures. The effect is bipolar in $V_\text{Gate}$ (not shown here) and is consistent with what has been reported for different materials~\cite{Paolucci2019c}. $V_\text{Gate}$ has little to no effect at low values until a sudden decrease close to $V_\text{Gate}^\text{C}$. At higher temperatures, the region where $V_\text{Gate}$ is ineffective widens, while $V_\text{Gate}^\text{C}$ is unaffected. In Fig.~\ref{fig:device}d, we show four IV curves for different $V_\text{Gate}$, taken at $T = 50$ mK. In line with previous field effect experiments, the retrapping current $I_\text{R}$ is not affected by $V_\text{Gate}$ until it coincides with $I_\text{C}$ (see also Appendix C). Above $V_\text{Gate}^\text{C}$ some residual non-linearity lingers, before the device becomes completely ohmic (see the 24 V line in Fig.~\ref{fig:device}d).~\cite{Paolucci2019c}

While the critical current is easily identified when the switch to the normal state is abrupt, this is less evident when $I_\text{C}$ is close to zero and the transition is more gradual. We have defined $I_\text{C}$ as the value of the bias current $I$ for which the differential resistance is larger than 10 $\Omega$, which is of the same order of magnitude as the normal state resistance $R_\text{N} \approx 25 \,\Omega$, and can be reliably identified over the background noise.
Unlike the switching process, the retrapping generally does not occur in one step, but tends to happen in two successive events (see e.g. the 18 V line in fig.~\ref{fig:device}d). The exact origin of this `partial' switching is not yet fully settled, but it is likely related to two local thermalization processes taking place in different regions of the device. 

The out-of-plane critical field versus temperature follows the phenomenological profile $B_\text{Z}(T)= B_\text{Z}(T=0) (1 - (T/T_\text{C})^2)$~\cite{Tinkham}, which yields $T_\text{C}=507$ mK and $B_\text{Z}(T=0) = 16.25$ mT (see Fig.~\ref{fig:BEverything}a and Appendix C). At $T=50$ mK, the critical field $B_\text{Y}^\text{C} \approx 850$ mT. Via the two critical magnetic fields, we estimate the London penetration depth $\lambda_\text{L}^\text{GL} \approx B_\text{Y}^\text{C} d / B_\text{Z}^\text{C} \sqrt{24} = 160$ nm~\cite{Tinkham} via the Ginzburg-Landau theory using d = 17 nm. Since the thickness of the SC film is $d \ll \lambda$, it is reasonable to assume that the in-plane field $B_\text{Y}$ penetrates the superconductor completely. Indeed, the critical in-plane field's temperature dependence $B_\text{Y}^\text{C} (T)$ is consistent with the evolution of a spin-split BCS condensate with a critical Zeeman field near the Clogston-Chandrasekhar limit $\mu_\text{B} H_\text{C} = \Delta_0 / \sqrt{2}$~\cite{Sarma1963,Adams2017,Bergeret2018} (see Fig. \ref{fig:BEverything}d and Appendix C). 

The simultaneous application of magnetic and electric fields is summarized in Fig.~\ref{fig:BEverything}, where the evolution of $I_\text{C}$ is plotted as a function of both temperature $T$ and out-of-plane magnetic field $B_\text{Z}$. The reduction of $I_\text{C}$ is monotonous in $B_\text{Z}$ and $T$.

\begin{figure*}[t]
\includegraphics[height=4.2cm]{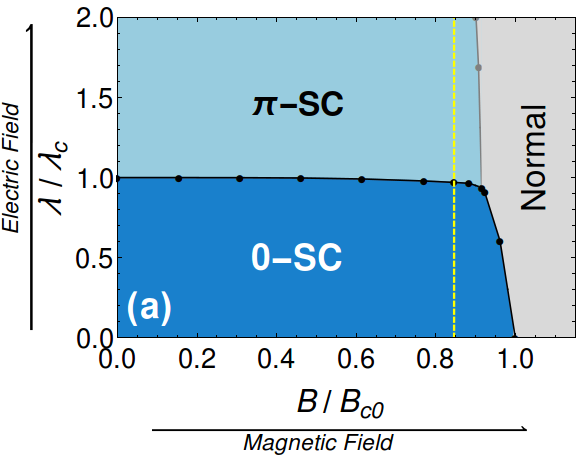} \hspace{0.4cm}
\includegraphics[height=4.1cm]{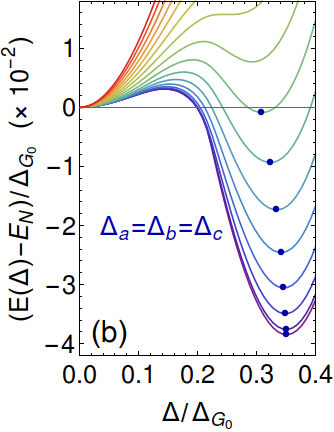}
\includegraphics[height=4.16cm]{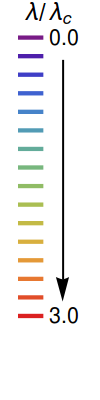}
\includegraphics[height=4.1cm]{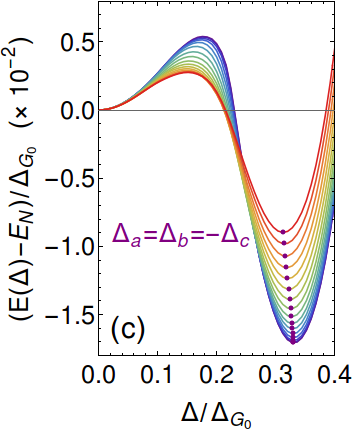}
\includegraphics[height=4.2cm]{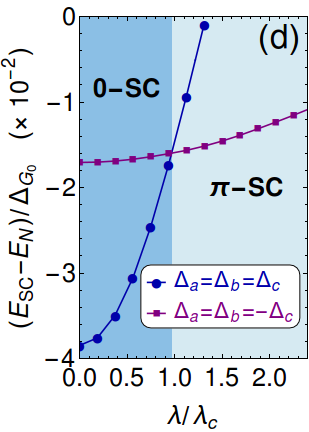}
\protect\caption{(a) Phase diagram in the $(B,\lambda)$ plane corresponding to a $B_Y$ applied Zeeman field and an effective electric field for an orbital Rashba coupling $\alpha_{OR}=0.2\,t$. We have three different phases: conventional superconducting state (0-SC), unconventional $\pi$-phase
($\pi$-SC), and normal metallic state (Normal). The transition line is obtained by comparing the free energy of the (0-SC) and ($\pi$-SC), as shown in panels (b-d). $B_{c0}$ is the critical field at $\alpha_{OR}=\lambda=0$. The critical $\lambda$ amplitude (or effective electric field) for the 0-$\pi$ transition does not change as a function of the applied magnetic field $B$, except close to the critical field $B_{c0}$. Other parameters: $n_z=6$ (number of layers); $t_\perp=1.5t, \mu=-0.4t, \eta=0.1$. (b-c) Behavior of the free-energy as a function of the SC-order parameter $\Delta$ for the conventional (b) and unconventional (c) SC states. $\Delta_{G_0}$ is the energy gap when $\alpha_{OR}=\lambda=0$ and $B=0$. The free energy is shown for several values of $\lambda$, ranging from 0 to $3\lambda_c$, for a fixed value of the magnetic field which is marked by the yellow line in the panel (a). The dark dots indicate the free energy minima. (d) Comparison of the minima of the free energy corresponding to the profiles in (b) and (c) for the 0 and $\pi$ phase, respectively. 
\label{0-pi-magfield}
%\\(if needed: in the scale of $t$  it is $B_{c0}=0.13t$ and
%for this value of $\alpha$ it is $\lambda_c \simeq 0.1 t$.)
} 
\end{figure*}

It is interesting to compare the dependence of $I_\text{C}$ on $T$ and $B_\text{Z}$, with the dependence on $V_\text{Gate}$ and $B_\text{Z}$, which is presented in Fig.~\ref{fig:BEverything}b. While the critical magnetic field $B_\text{Z}^\text{C}$ decreases continuously with temperature, the same is not true for $V_\text{Gate}$. For $V_\text{Gate} < 17$ V, the dependence of $I_\text{C}$ on $B_\text{Z}$ is unaffected. Only when $V_\text{Gate}$ exceeds this value, we see a reduction in both $I_\text{C}$ and a sharp decrease of $B_\text{Z}^\text{C}$. The dependence of $B_\text{Z}^\text{C}$ on $V_\text{Gate}$ is shown in Fig.~\ref{fig:BEverything}c, for three different temperatures. Even at higher $T$, the onset of the reduction of $B_\text{Z}^\text{C}$ is not significantly changed.

The complete evolution of $I_\text{C}$ as a function of both $T$ and $B_\text{Y}$ is shown in Fig.~\ref{fig:BEverything}d. Analogous to the effect of $B_\text{Z}$, $I_\text{C}$ is reduced monotonously. Also the behavior of $I_\text{C}$ versus $T$ and $V_\text{Gate}$ is similar; for $V_\text{Gate} < 17$ V, the dependence of $I_\text{C}$ on $B_\text{Y}$ is not significantly affected. Figure~\ref{fig:BEverything}f depicts the evolution of the critical magnetic field $B_\text{Y}^\text{C}$ versus $V_\text{Gate}$ for several $T$.
For neither $B_\text{Z}$ nor $B_\text{Y}$ does the relation between $I_\text{C}$, $B$ and $V_\text{Gate}$ depend on the sign of either $B$ or $V_\text{Gate}$.

\subsection{Theoretical phase diagram}
Starting from the zero magnetic field configuration, the coupling $\lambda$ (i.e. the electric field) can drive transitions of the type 0-$\pi$ (i.e. conventional-to-unconventional superconducting phase) or superconducting-normal depending on whether the $\alpha_{OR}$ coupling is smaller or comparable to the planar kinetic energy scale set by the hopping amplitude $t$~\cite{mercaldo2019}. Here, the $\pi$-phase means that the SC order parameter in a given band has a different sign with respect to that in the other bands contributing to the pairing at the Fermi level. Instead, in the 0-phase there is no phase difference among the bands. Within our modeling the three bands contributing at the Fermi level (i.e., $a,b,c$) are coupled through OR effects; in the $\pi$-phase we have that $\Delta_a=\Delta_b=-\Delta_c$. 

The essential outcome of our modelling is that the electric field is able to break the inter-orbital phase rigidity before fully suppressing the amplitude of the order parameter. This is a consequence of the inversion-symmetry breaking at the surface layers induced by the external electric field which polarizes the orbitals of the electronic states at the Fermi level. The electric field in this context has two main consequences for the phase diagram. It rearranges the orbital dependent superconducting phases with a $\pi$-shift ($\pi$-phase) and it suppresses the amplitude of the order parameter by increasing the population of depaired orbitally polarized quasi-particles (electrically driven normal phase). Both phases are marked by a vanishing supercurrent, however, the underlying mechanisms that leads to the supercurrent suppression is fundamentally different. In the $\pi$-phase the vanishing supercurrent is due to orbitally driven frustration of the phase of the superconducting order parameter, while in the electrically induced normal phase it is due to the suppression of the pairing order parameter. These two scenarios can be distinguished by the response of the critical voltage to an external magnetic field.

Let us start by considering the behavior of the $\pi$-phase from a representative case with $\alpha_{OR}=0.2~t$ at zero temperature (Fig. \ref{0-pi-magfield}). For $B=0$ the superconductor undergoes a 0-$\pi$ transition above a critical $\lambda$ which is proportional to the applied electric field. As expected, when considering a non-vanishing Zeeman field $B$, the superconductor exhibits a transition into a normal state if $B$ exceeds a critical field $B_\text{C}$. This SC-Normal transition is also obtained in the presence of a non-vanishing $\lambda$. Remarkably, both the $0$-$\pi$ phase boundary and the critical lines separating the 0- or $\pi$-phases from the normal state shows a weak interplay between the electric and magnetic fields (Fig. \ref{0-pi-magfield}(a)). Indeed, $\lambda_c$ does not exhibit significant changes as a function of the magnetic field $B$, except for close to the transition point. A similar behavior is also observed for $B_\text{C}$. The phase diagram is determined by evaluating the behavior of the free energy at a given magnetic field for the $0-$ and $\pi-$ phases (Fig. \ref{0-pi-magfield}(b)-(c)). While the free energy minimum of the $0$-phase is strongly affected by the electric field, via $\lambda$, the $\pi$-phase is more resilient and at $\lambda \sim \lambda_c$ there is a transition from 0- to $\pi$-phase due to the crossing of the corresponding free energies (Fig. \ref{0-pi-magfield}(d)). This transition is starkly unaffected by the magnetic field $B$ and it varies only close to the critical point where both 0- and $\pi-$ phases can be brought into the normal state. The weak dependence of $\lambda_\text{C}$ on the magnetic field can be ascribed to the character of the $\pi$-phase, marked by only a rearrangement of the relative phases between the band-dependent SC order parameters, while their amplitudes do not significantly vary across the transition. 

%%%%%
\begin{figure*}[bt]
\includegraphics[width=0.3\textwidth]{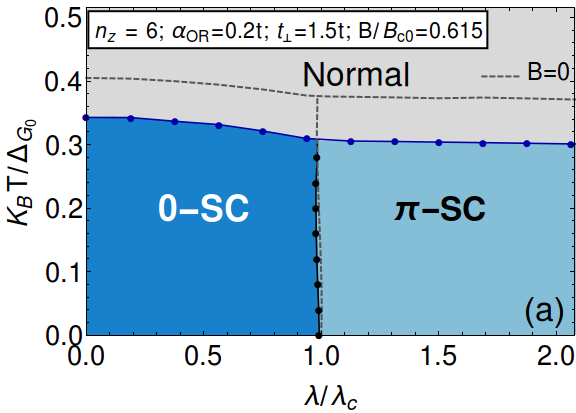} \hspace{0.5cm}
\includegraphics[width=0.32\textwidth]{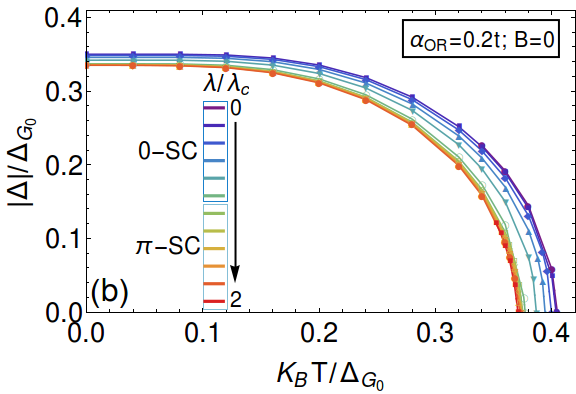}
\includegraphics[width=0.32\textwidth]{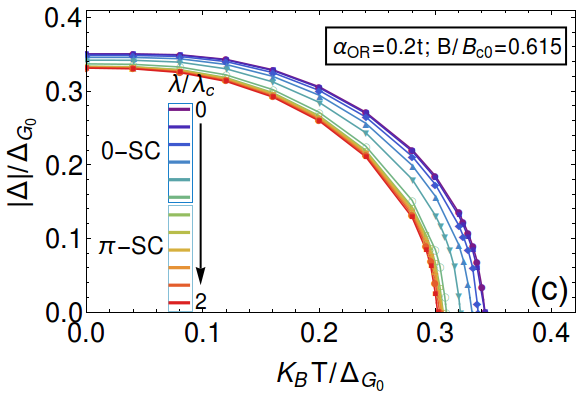}
\protect\caption{(a) Phase diagram in the $(\lambda,T)$ plane showing three different states: conventional superconducting state (0-SC), unconventional
($\pi$-SC), and normal state for $\alpha_{OR}=0.2t$ for $B = 0.615 B_{c0}$.  We assume an in-plane magnetic field orientation, e.g. $B_y$. The critical $\lambda$ amplitude for the 0-$\pi$ transition (black line) does not change as a function of temperature. The transition from the SC to normal state (blue line) is of second order. The gray dashed lines are the transition lines in absence of magnetic field $(B=0)$. (b-c) Behavior of the order parameter as a function of temperature for $B=0$ (panel (b)) and $B\neq0$ (panel(c)) and for several values of $\lambda$. } 
\label{fig4}
\end{figure*}
%%%%%

Now, one can try to compare the results of Fig. 2(f) with those of Fig. 3(a). We observe that in Fig. 3(a) the critical magnetic field B$_\text{C}$, setting the boundary between the 0- or $\pi$-phase and the Normal state, is practically unaffected by the variation of the electric field. This outcome can capture the behavior of the experimental critical field in Fig. 2(f) that is also flat and is not significantly varying except close to the critical voltage. We argue that the $\pi$-phase is the configuration that can be induced by the electric field already before reaching the critical voltage to account for the decrease of the supercurrent. Within the $\pi$-phase the supercurrent suppression is firstly driven by a pure phase mechanism due to the inter-band sign frustration and then further amplified by the reduction of the amplitude of the superconducting order parameter. On the other hand, the rapid decrease of the critical magnetic field close to the critical voltage is due to the occurrence of normal state configurations. In this regime, we expect that the magnetic field phenomenology can be captured by the character of the 0-normal phase transition.

When considering the transition from the 0-SC to N state by varying the electric field amplitude ($\lambda$) at a larger value of the $\alpha_{OR}$ coupling, one finds a stronger correlation between the critical electric field and the magnetic field (see Fig. 7 in the Appendix B). 

%{\color{blue}{This behavior can also be qualitatively understood by considering that within the employed model the electric field couples directly to the orbital degrees of freedom and the electrically driven superconducting phases are marked by orbital dephasing and orbital depairing processes. On the other hand, the magnetic Zeeman field mainly couples to the spin and thus leads to a breakdown of the superconducting phase by unpairing the spin of the Cooper pairs. Hence, while the different electronic impact of the electric and magnetic fields on the superconductor prefigures a weak interrelation between the critical voltage and critical magnetic fields, the orbital dephasing is remarkably more resilient to the application of a magnetic field than the orbital depairing one.}}

%Furthermore, we have also investigated the evolution of the phase diagram in temperature. The 0-$\pi$ phase boundary is unaffected by the temperature, indicating that the electric field threshold is not altered by thermal effects until reaching the critical superconducting temperature.

\begin{figure}[ht]
\includegraphics[height=9cm]{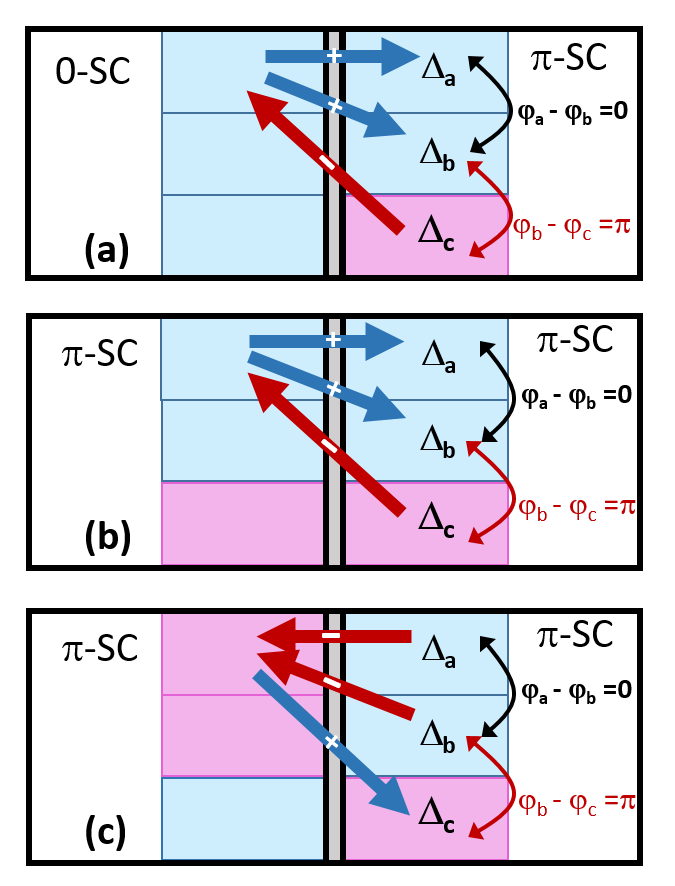}
\protect\caption{(a) Schematic description of the competing directions among Josepshon pair currents (arrows) at the interface between superconducting domains having a multiband character with 0- and $\pi$-phase (a) or with $\pi$-phases on both sides (b)-(c). For graphical clarity we have depicted only the arrows from the $a$ bands to all the other bands across the interface. Taking into account the charge transfer processes at the interface of the superconducting domains between homologue or different bands there can be Josephson currents with positive or negative sign. The resulting outcome is to have an overall tendency to cancel out the total supercurrent. For the interface between $\pi$-phase domains, since the configurations (b) and (c) are approximately degenerate in energy, it is plausbile to expect an enhanced suppression of the supercurrent especially when considering an inhomogeneous superconducting films (e.g. polycrystalline). 
%in a way that can be less dependent of the interface electronic processes. 
\label{fig5} } 
\end{figure}

In order to assess the role of the thermal fluctuations we have also determined the phase diagram at finite temperature for the case of small orbital Rashba coupling.
In Fig. \ref{fig4} we report the phase diagram with the evolution of the transition lines among the $0$-, $\pi$-SC phases and the normal metallic state by considering the effects of the temperature and of the effective electric field through the $\lambda$ coupling. We compare the zero magnetic field case with one representative configuration corresponding to $B\sim 0.6 B_c$. There are two relevant observations to highlight: firstly, the critical boundary from the 0-SC state to the $\pi$-phase is substantially unaffected by the temperature and by the applied magnetic field. Secondly, the critical temperature for the superconducting-normal transition is also independent on the 0- or $\pi$-character of the superconducting phase, as is observed experimentally. 
The evolution of the superconducting order parameters in temperature demonstrate a conventional trend with a weak dependence on the electric (via $\lambda$) and magnetic fields ($B$) as explicitly reported in Fig. \ref{fig4}(b),(c).

\section{Discussion and conclusions}
Comparing the above theoretical results with the experimental observations we argue that the experimental outcome supports the $\pi$-phase for accounting the phenomenology of the magnetic field response of the SC nano-bridges in the presence of an electrostatic gating. We note that in the $\pi$-phase, the presence of inter-band $\pi$-phase slips can naturally account for a suppression of the supercurrent, due to a cancellation between positive and negative pair currents among the various bands that are present at the Fermi level. 
In Fig. \ref{fig5} we schematically depict this scenario assuming that the superconducting film can form domains both due to the expected inhomogeneous distribution of the electric field on the surface and also due to the intrinsic polycrystalline character of the investigated materials. Then, before the electric field is able to fully suppress the superconducting state and driving it into a normal metal configuration it is plausible to expect that an inhomogeneous phase with 0-$\pi$ (small electric fields) and $\pi$-$\pi$ interfaces (with increasing electric field) is achieved. 
Moreover, since the $\pi$-phase does not exhibit spatial modulations or gradients of the superconducting order parameter, we expect a weak influence from the formation of a vortex phase, as induced by the out of plane magnetic field $B_\text{Z}$.
Thus, this supports the observation that the electric field is able to disrupt the superconducting state by primarily inducing $\pi$-phase slips between the electronic states that contribute to the pairing at the Fermi level. This remark is also consistent with the enhancement of non-thermal phase fluctuations that have been observed in the switching current distributions of Ti Dayem bridges.\cite{Puglia2019} 

%\section{Summary} 
In conclusion, we have investigated the suppression of supercurrent effected by the electric field, combined with and in-plane, or out-of-plane magnetic field, and ascertained that the two are weakly coupled: the critical magnetic fields are only affected for gate voltages close to the critical gate voltage. These findings are consistent with a microscopic model based on a multiband description of the superconducting state where the electric field is assumed to induce an electrostatic interaction at the surface and in turn a strong orbital polarization at the Fermi level. The effect of a magnetic field on the electrically driven phase transitions has been thoroughly explored and the way the electric and magnetic fields can affect the superconductivity in thin films has been set out clearly. 
Furthermore, we have, for the first time, realized a complete suppression of the critical current in an aluminium-based Dayem bridge via electrostatic gating. Since aluminium is an important material from the technological point of view, this paves the way for future applications of the electric field effect.

For completeness, it is also valuable to comment on recent results showing an increase of quasi-particle population induced by gate effects on superconducting nanowires\cite{Alegria2020}. Our proposed model, although completely different in microscopic structure and nature in comparison to the high-energy injection scenario, is however compatible with the increase of the quasi-particle population and the modification of the in-gap spectral weight. In fact, in both the 0-phase and especially in the $\pi$-phase (due to the sign frustration of the superconducting order parameters) the electric field is able to induce a variation of the in-gap quasi-particles through the orbital polarization effect.

%Furthermore, we have to mention that in a very recent experiment \cite{suspended} a fully suspended gate-controlled Ti nano-transistor has been realized where a complete suppression of the critical supercurrent has been observed. The experiment was specifically designed to eliminate any direct injection of quasi-particles through the substrate thereby leaving cold electron field emission through the vacuum the only possible channel of charge injection into the superconductor. The observed features have been further modeled and put forward as evidence of the absence of thermal effects or cold electron field emission in the gate-controlled superconducting phases. Although, these recent experimental results seems to rule out charge injection as possible mechanism, we believe that the nature of the electric gating effect in all metallic superconductors is still an open challenge that requires further investigation to fully solve the puzzle.}}

Although, recent experimental results on suspended Ti nanowires\cite{suspended} seem to rule out charge injection through the substrate surface as a possible mechanism, we believe that the nature of the electric gating effect in all metallic superconductors is still an open challenge that requires further investigation to fully solve this exciting puzzle.

\begin{acknowledgments}
E.S. and F.G were partially supported by EU’s Horizon 2020 research and innovation program under Grant Agreement No. 800923 (SUPERTED).
\end{acknowledgments}

\appendix
%\section{Appendix}

\section{Model derivation: interactions induced by surface electric field}

The external electric field on the surface of the superconductor is parallel to the $\hat{z}$-direction and can be described by a potential $V_{s}=-E_{s} z$ with $E_s$ being constant in amplitude (assuming the electric charge $e$ is unit). We consider a Bloch state representation and explicitly evaluate the matrix elements of the electrostatic potential $V_s$. 
Since the translational symmetry is broken along the $\hat{z}$-direction due to the finite thickness of the thin film and for the electric field, the out-of-plane momentum is not a good quantum number. Thus, a representation with a Bloch wave function associated to each layer is suitable to evaluate the effects of the electric field and the way it enters in the tight-binding modelling.
Here, we use the index $i_z$ to label different Bloch wave functions along the $\hat{z}$-direction as follows
\begin{eqnarray}
\psi_{{\bf k},\beta}({\bf r},i_z) =\frac{1}{\sqrt{N}}\sum_{\nu} \exp[i {\bf k} \cdot {\bf R}_{\nu, i_z}] \phi_{\beta}({\bf r}-{\bf R}_{\nu,i_z})
\end{eqnarray}
\noindent with the Bravais vector ${\bf R}_{\nu,i_z}$ identifying the position of the atoms in the $x-y$ plane for the layer labelled by $i_z$, $\beta$ indicating the atomic Wannier orbitals, and $N$ the total number of atomic sites. A central aspect in the derivation is that the atomic Wannier functions span a manifold with non-vanishing angular momentum ${\bf L}$. 
To proceed further, we demonstrate how orbitally driven Rashba-like splitting occur for $p$- states $\{p_x,p_y,p_z\}$ due to the presence of the inversion symmetry breaking potential $V_s$ by evaluating the corresponding matrix elements in the Bloch basis. Similar results can be obtained following the same approach for hybridized $sp$- and $d$-states.
\\
The $p$-orbitals for a given atomic position ${\bf R}_{\nu,i_z}$ are expressed as
\begin{eqnarray}
&&\phi_{x}({\mathbf{r}})=f(r)\,x\, \exp\left[-\frac{Z r}{n a_M}\right] \nonumber \\ 
&& \phi_{y}(\mathbf{r})=f(r)\,y\, \exp\left[-\frac{Z r}{n a_M}\right] \nonumber \\
&& \phi_{z}(\mathbf{r})= f(r)\,z\, \exp\left[-\frac{Z r}{n a_M}\right]
\end{eqnarray} 
\noindent with $f(r)=f_0(Z,n)\,L_{n+1}^{3}(t)$, $f_0(Z,n)$ being a numerical prefactor, $Z$ the atomic number, $n$ the principal quantum number, $t=2 Z r/(n a_M)$, $a_M=a_0(1+m_e/M)$ with $a_0$ the Bohr radius, $m_e$ and $M$ the mass of the electron and nucleus, and $L_p^q(t)$ the associated Laguerre polynomials. These $p$-orbitals are also linked with the eigenstates $\{|0\rangle, |1\rangle,|\bar{1}\rangle \}$ of the $L_z$
component of $L=1$ angular momentum with quantum numbers $\{0,1,-1\}$.
As done in the main text, $(a,b,c)$ will be used to indicate
the $p$-orbitals. 
\begin{figure*}[bt]
\includegraphics[height=4.cm]{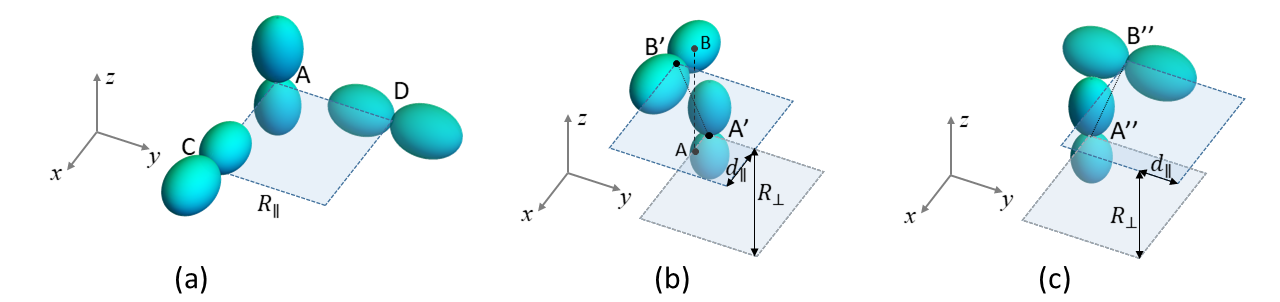}
\protect\caption{Schematic figure describing the atomic positions for the determination of the electrostatic energy associated to the intra- and inter-layer electronic processes for the $p$-orbitals. (a) sketch of the nearest neighbor atomic positions along the $(x,y,z)$ symmetry directions. (b) and (c) describe schematically the in-plane displacements that are related with the inter-layer orbital Rashba coupling. The displayed orbitals are those that contribute in the in- and out-of-plane electronic processes for the electric field induced interactions.}
\label{fig:schema}
\end{figure*}
Now, in order to evaluate the consequence of the electrostatic potential, we need to determine the matrix elements in the Bloch state representation within the same layer and in the neighbors layers along the $\hat{z}$-direction. These terms will provide, in turn, the amplitude of the orbital Rashba coupling $\alpha_{OR}$ and $\lambda$, respectively.
Let us start by calculating the intra-layer interaction 
\begin{eqnarray}
A^{||}_{l,m}=&& c_{\psi} \langle \psi_{{\bf k},l}({\bf r},i_z) | (-E_s z) | \psi_{{\bf k},m}({\bf r},i_z) \rangle \nonumber \\
=&&  c_{\psi} (-E_s) \frac{1}{N} \sum_{\nu,\gamma} \exp[i {\bf k} \cdot \left({\bf R}_{\nu, i_z}-{\bf R}_{\gamma, i_z}\right)] \times \nonumber \\
&& \times \int d^3 {\bf{r}} \phi^{*}_{l}({\bf r}-{\bf R}_{\nu,i_z})\,z\, \phi_{m}({\bf r}-{\bf R}_{\gamma,i_z}) \,
\end{eqnarray}
\noindent with $l$ and $m$ spanning the orbital space, and $c_{\psi}$ the normalization factor of the Bloch state.
Since the functions $\phi_{l}({\bf r}-{\bf R}_{\gamma,i_z})$ are localized around each atomic position one can restrict the summation to leading terms which are those corresponding to the same site, i.e. ${\bf R}_{\nu, i_z}={\bf R}_{\gamma, i_z}$, and to nearest-neighbor sites, i.e. ${\bf R}_{\nu, i_z}={\bf R}_{\gamma, i_z}\pm {\bf a}_{x,y}$, with ${\bf a}_{x,y}$ being the connecting vectors of nearest-neighbor atoms in the $x-y$ plane. The term for ${\bf R}_{\nu, i_z}={\bf R}_{\gamma, i_z}$ is zero due to the odd-parity symmetry of the atomic functions. 
Then, assuming that the distance between two in-plane nearest-neighbor atoms is $R_{||}$, the amplitude $A^{||}$ can be expressed in a matrix form as 
\begin{eqnarray}
{\hat{A}}^{||}=c_{\psi} (-E_s)\,R_{||}\,I_{||}(R_{||};Z,n) \left[\sin(k_x R_{||}) L_{y} -\sin(k_y R_{||}) L_x \right] \nonumber \\
\label{Ainplane}
\end{eqnarray}
\noindent with $I_{||}(R_{||};Z,n)$ being a function of the relative atomic distance $R_{||}$, the atomic number $Z$ and the principal quantum number of the Wannier functions $n$, respectively.
Hence, comparing $A^{||}$ with the term of the Hamiltonian associated with the orbital Rashba coupling, we have that the strength of the orbital Rashba coupling $\alpha_{OR}$ is expressed as
\begin{eqnarray}
\alpha_{OR}=(-E_s)\,R_{||}\,I_{||}(R;Z,n) c_{\psi}
\end{eqnarray}
and it is proportional to the intensity of the applied electric field $E_s$ and to the amplitude $I_{||}(R_{||};Z,n)$. The form of ${\hat{A}}^{||}$ in Eq. \ref{Ainplane} is due to the structure of the expectation values of the electrostatic potential between neighbors Wannier functions. 
If we consider schematically the atomic positions $P_A=[0,0,-\frac{R_{\perp}}{2}]$, $P_B=[0,0,\frac{R_{\perp}}{2}]$, $P_C=[R_{||},0,-\frac{R_{\perp}}{2}]$, $P_D=[0,R_{||},-\frac{R_{\perp}}{2}]$, for a cubic geometry in Fig. \ref{fig:schema}(a), we have that
\begin{eqnarray}
\langle\phi_{A,m}| E_s z| \phi_{C,m}\rangle &=&0 \quad {\text{for}}\; m=a,b,c \\
\langle\phi_{A,a}| E_s z| \phi_{C,b} \rangle &=& \langle\phi_{A,b}| E_s z| \phi_{C,c}\rangle =0 \\
\langle\phi_{A,a}| E_s z| \phi_{C,c}\rangle &=& -E_s R_{||} I_{||}(R_{||};Z,n)\\
\langle\phi_{A,c}| E_s z| \phi_{C,a}\rangle &=& +E_s R_{||} I_{||}(R_{||};Z,n) \,.
\end{eqnarray}
\\
The same expressions are obtained along the $\hat{y}$-direction for the orbitals $b$ and $c$. 
In a similar way, one can proceed for the matrix elements of the electrostatic potential between Bloch states in adjacent layers expressed as
\begin{eqnarray}
A^{\perp}_{p,q}=&&c_{\psi} \langle \psi_{{\bf k},p}({\bf r},i_z) | (-E_s z) | \psi_{{\bf k},q}({\bf r},i_z\pm 1)\rangle \,.
\label{Aoutplane}
\end{eqnarray}
As for the in-plane amplitude, one can expand the summation over all the Bravais lattice. However, in this case there are contributions which are non-vanishing for ${\bf R}_{\nu, i_z}={\bf R}_{\gamma, i_z\pm 1}$ and, thus, we focus on these contributions
\begin{eqnarray}
A^{\perp}_{p,q}=c_{\psi} (-E_s)
\int d^3 {\bf{r}} \phi^{*}_{p}({\bf r}-{\bf R}_{\nu,i_z})\,z\, \phi_{q}({\bf r}-{\bf R}_{\nu,i_z\pm 1}) \,. \nonumber \\
\end{eqnarray}
To proceed further we notice that the amplitude $A^{\perp}_{p,q}$ is in general complex because the electric field induces a time dependent vector potential along the $\hat{z}$-direction that affects the relative phase of the Bloch functions in neighbor layers. This implies that one cannot fix the gauge in a way that the Bloch states in adjacent layers at the surface, e.g. $\psi_{{\bf k},p}({\bf r},i_z=1)$ and $\psi_{{\bf k},p}({\bf r},i_z=2)$, have the same phase. This is an overall phase factor that does not influence the amplitude of the term $A^{\perp}_{p,q}$. Below, we proceed by considering the contribution which leads to a coupling between the electric field and the orbital polarization. 
The form of ${{A}}^{\perp}$ is due to the strucure of the matrix elements of the electrostatic potential between Wannier functions in neighbor layers along the $\hat{z}$-direction.
Hence, one has to evaluate the following integrals 
\begin{eqnarray}
\int d^3 {\bf{r}} \phi^{*}_{p}({\bf r}-{\bf R}_{\nu,i_z})\,z\, \phi_{q}({\bf r}-{\bf R}_{\nu,i_z\pm 1}) \,.
\end{eqnarray} 
\noindent for nearest neighbor atoms along the $\hat{z}$-direction as schematically shown in Fig. \ref{fig:schema}.

For the inter-layer term, it turns out that the electric field can induce an orbital polarization on nearest neighbors atoms only if one allows for displacements/distortions of the atoms in the plane with respect to the high-symmetry positions. This physical scenario is  sketched in Fig. \ref{fig:schema}(b,c).
The analysis is performed by considering the following positions for the atoms $A'$ and $B'$ in the plane, $P_{A'}=[-\frac{d_{||}}{2},0,-\frac{R_{\perp}}{2}]$, $P_{B'}=[\frac{d_{||}}{2},0,\frac{R_{\perp}}{2}]$. 
As for the intra-plane case, we have that the relevant non-vanishing integrals are those related to the $L_x$ and $L_y$ components of the angular momentum, namely we have the $L_y$ component that is active for an atomic displacement along the $\hat{x}$-direction. Within a first order expansion in $d_{||}/R_{\perp}$ one obtains
\begin{eqnarray}
\langle \phi_{A',a} | E_s z |\phi_{B',a} \rangle &=& \langle \phi_{A',b} | E_s z |\phi_{B',b} \rangle = \langle \phi_{A',c} | E_s z |\phi_{B',c} \rangle = 0 \nonumber \\
\langle \phi_{A',a} | E_s z |\phi_{B',b} \rangle &=& 0 \nonumber \\ 
\langle \phi_{A',c} | E_s z |\phi_{B',a} \rangle &=& - \langle \phi_{A',a} | E_s z |\phi_{B',c}\rangle = E_s d_{||} I_{\perp}(R_{\perp};Z,n) \nonumber \\
\langle \psi_{A',b} | E_s z |\psi_{B',c} \rangle &=& 0 \,.
\end{eqnarray}
\\
A similar analysis for a distortive mode along the $\hat{y}$-direction would give a non-vanishing amplitude only for the wave functions $\phi_{c}$ and $\phi_{b}$. Assuming that the atomic distorsions along the $\hat{x}$- and $\hat{y}$-directions have the same amplitude (Fig. \ref{fig:schema}(c)), the resulting expression for the matrix $\hat{A}^{\perp}$ is 
\begin{eqnarray}
\hat{A}^{\perp}=E_s d_{||} I_{\perp}(R_{\perp};Z,n) c_{\psi} (L_x+ L_y) \,.
\end{eqnarray}
\noindent Hence, comparing the structure of $\hat{A}^{\perp}$ with the inter-layer asymmetric interaction introduced in the Hamiltonian, we have that 
\begin{eqnarray}
\lambda=E_s d_{||} I_{\perp}(R_{\perp};Z,n) c_{\psi} \,.
\end{eqnarray}

\begin{figure*}[bt!]
\includegraphics[height=4.12cm]{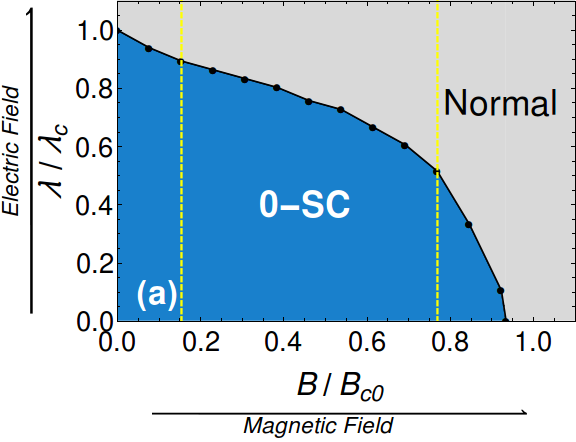} \hspace{0.4cm}
\includegraphics[height=4.12cm]{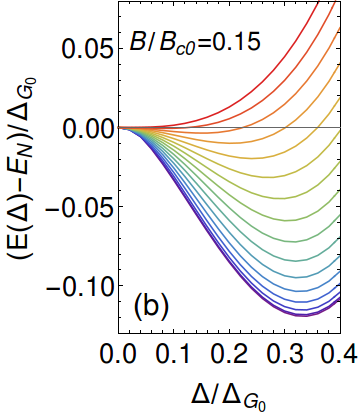}
\includegraphics[height=4.12cm]{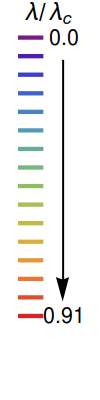}
\includegraphics[height=4.12cm]{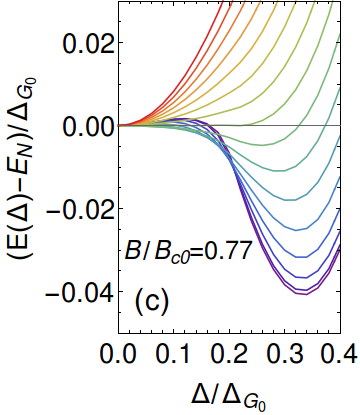}
\includegraphics[height=4.12cm]{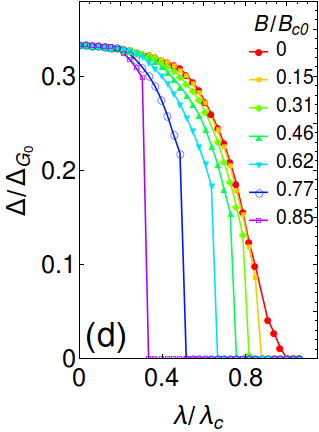}
\protect\caption{(a) Phase diagram in the $(B,\lambda)$ plane  for $\alpha_{OR}=1.0t\,$ assuming an in-plane magnetic field (e.g. $B_Y$). For this value of $\alpha_{OR}$, there is no stable ($\pi$-SC) state. Indeed the free-energy for the $\pi$-SC case has a higher value with respect to the normal phase. $B_{c0}$ is the critical field when $\alpha_{OR}=\lambda=0$ and the other parameters are: $n_z=6$ (number of layers); $t_\perp=1.5\,t, \mu=-0.4\,t, \eta=0.1$. (b-c) Behavior of the free-energy as function of the SC-order parameter $\Delta$ for two  different values of the magnetic field $B$ (marked by the yellow lines in (a)). $\Delta_ {G_0}$ is the energy gap when $\alpha_{OR}=\lambda=0$ and $B=0$. The free energy is shown for several values of $\lambda$, ranging from 0 to $\sim 0.9\lambda_c$ ($\lambda_c \simeq 0.33 t$). (d) Behavior of the SC order parameter $\Delta$ as a funciton of $\lambda$ for different applied magnetic field.
%When $B=0$ the transition is smooth (second order), while it is first order in the presence of magnetic field. %The jump in the OP increases while increasing the  the field $B$. 
%\\(if needed: in the scale of $t$  
%for this value of $\alpha$ it is $\lambda_c \simeq 0.33 t$.)
%
} 
\label{sup:fig1}
\end{figure*}

\section{Superconducting-normal phase boundary: role of magnetic field}

In the main text we have shown that for weak orbital-Rashba couplings compared to $t$, i.e. the in-plane kinetic energy scale, the increase of the inter-layer interaction $\lambda$ can drive a rearrangement of the interband superconducting phase difference resulting into a $\pi$-phase above a critical threshold for $\lambda$. We have also demonstrated that this transition is substantially unaffected by the presence of a magnetic field. 
In this Appendix, for completeness we also investigate another interesting regime that refers to values of the orbital Rashba coupling that are comparable to $t$, where the $\pi$-phase is not stable and the variation of the ampltidue of the inter-layer asymmetric coupling $\lambda$ drives a transition from the superconducting to the normal metal. 
In Fig. \ref{sup:fig1}(a) we observe that at zero magnetic field, for $\alpha_{OR}=1.0\,t$, the superconducting state with uniform orbital phase (0-SC) can be tuned into a normal state by increasing the amplitude of $\lambda$ above a critical amplitude $\lambda_c$. The type of transition is continuous as one can notice by inspection of the superconducting order parameter (Fig. \ref{sup:fig1}(d)). Here, we recall that the amplitude of $\lambda$ measures the strength of the applied electric field. 
The evolution of the critical line separating the 0-SC state from the normal metallic state indicates that one can destroy the superconducting phase with a smaller amplitute of the $\lambda$ coupling in the presence of an applied magnetic field. In particular, close to the critical magnetic field the strength of the $\lambda$ coupling can be tuned to be vanishingly small. 
This would imply that the threshold of the electric field to disrupt the superconductivity can be tuned to zero by the simultaneous presence of an applied magnetic field. In this respect, the behavior of the critical line is not compatible with the experimental observation that the electric field amplitude to disrupt the superconducting phase is weakly dependent on the strength of the applied magnetic field.
It is reasonable to expect such behavior for the 0-SC/normal transition line because the magnetic field acts as a source of spin pair-breaking and thus it tends to reduce the energy of the superconducting phase and in turn it favors the stability of the normal state.
We point out that the transition line is first order type and there is net jump at the boundary of the superconducting order parameter which grows with the increase of the magnetic field strength.
We have also verified that a change in the number of layers does not alter the outcomes of the results by explicitly evaluating the case with $n_z=12$. 

\section{Critical magnetic fields versus temperature and retrapping currents}
\begin{figure}
\centering
\includegraphics[width=\columnwidth]{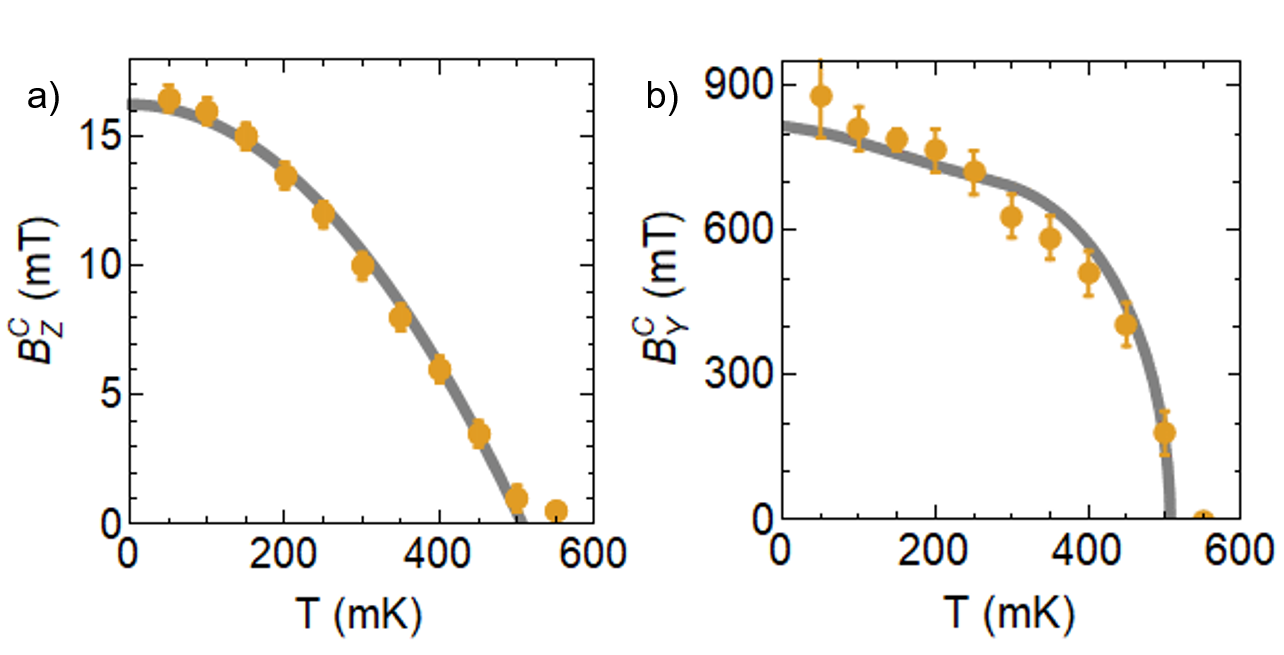}
\caption{a) Critical out-of-plain magnetic field $B_\text{Z}^{\text{C}}$ versus temperature, error bars indicate the resolution in B. Fitted with an empirical expression (see main text).
b) Critical in-plain magnetic field $B_\text{Y}^{\text{C}}$ (along the direction of the current) versus temperature. Error bars indicate the resolution in B. Fitted with the calculated temperature dependence of the critical field assuming perfect spin paramagnetism (see main text).
}
\label{fig:BcritvT}
\end{figure}
\begin{figure*}
\centering
\includegraphics[width=\textwidth]{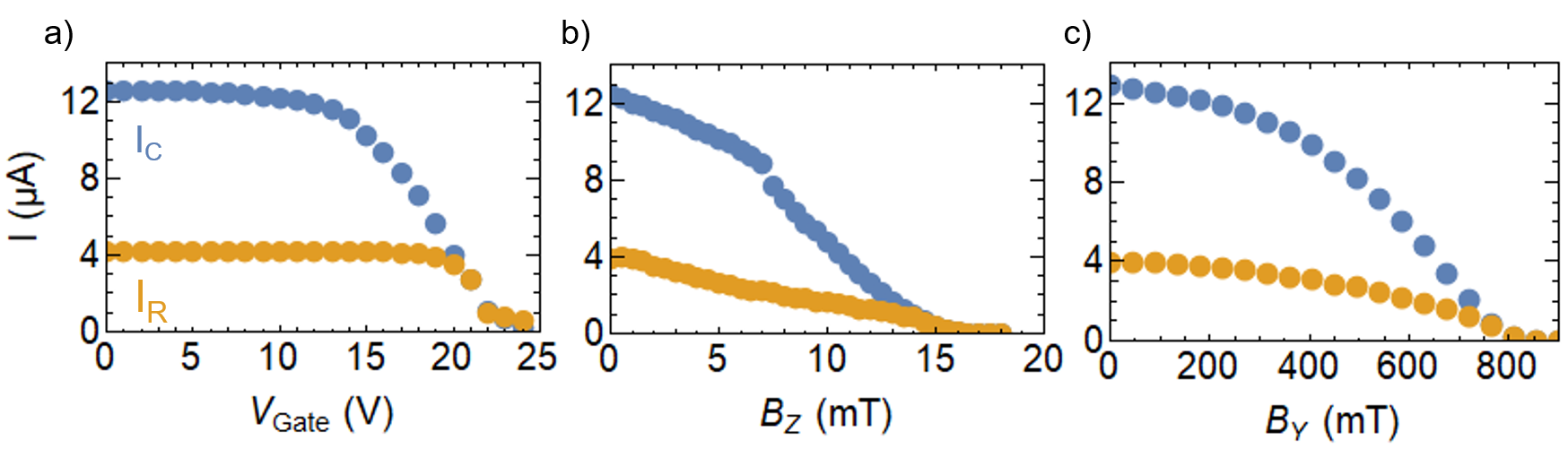}
\caption{a) The critical and retrapping current versus gate voltage, b) out-of-plane field $B_\text{Z}$ and c) in-plane field $B_\text{Y}$.}
\label{fig:SwitchingAndRetrapping}
\end{figure*}

It is well known, that magnetic fields suppress superconductivity, although the in- and out-of-plane magnetic fields do so in very different ways. In Fig.~\ref{fig:BcritvT}a, we show the critical out-of-plane magnetic field versus temperature, where the error bars indicate the resolution in $B_\text{Z}$. The data agrees well with the phenomenological expression $B_\text{Z}(T)= B_\text{Z}(T=0) (1 - (T/T_\text{C})^2)$~\cite{Tinkham}, which yields a critical temperature $T_\text{C}=507$ mK and critical out-of-plane field of $B_\text{Z}(T=0) = 16.25$ mT.

Fig.~\ref{fig:BcritvT}b, shows the critical in-plane field $B_\text{Y}^\text{C} (T)$, fitted with a calculation of the temperature dependence of the critical in-plane field assuming a homogenous spin splitting, while minimizing the free energy.~\cite{Sarma1963,Adams2017,Bergeret2018}

%\section{Critical and retrapping current dependencies}

A noteworthy aspect of the electric field effect is that the the electric field seems to become less effective at higher temperatures, i.e. a higher voltage is required before the critical current is reduced, while the critical gate voltage where the supercurrent is zero is not affected. Similarly it can be interesting to consider the retrapping current, and how it responds to the electric field as opposed to the magnetic fields. Unlike the dependence of the retrapping current on the temperature (see Fig.~\ref{fig:device}b), the retrapping current seems to be completely unaffected by $V_\text{Gate}$, until the critical current is reduced to the original value of the retrapping, see Fig.~\ref{fig:SwitchingAndRetrapping}. From that point on, the critical and retrapping current coincide. On the other hand, for the magnetic fields we see that both $I_\text{C}$ and $I_\text{R}$ are immediately affected, a reflection of the suppression of the superconducting order parameter the magnetic fields impose.

%\bibliographystyle{apsrev4-1}
%\bibliography{AL_DB_bib}

\end{document}